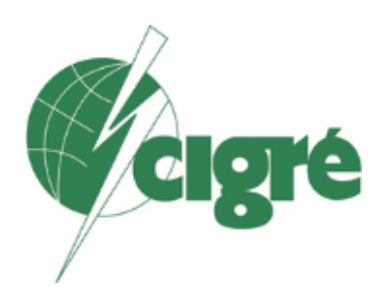

# Long-Term Operational Planning Using Scenario-Based System Load Forecasting

**A DEBNATH, S S GHOSH, J DE LA REE, A FREEMAN, K D JONES**
**Dominion Energy**
**United States of America**

## SUMMARY

The rapid expansion of hyperscale data centers (Large Loads), driven by artificial intelligence and digital transformation, is significantly increasing electricity demand in Northern Virginia. Dominion Energy, the region's primary electric utility, plays a critical role in supporting this growth, which necessitates substantial transmission infrastructure reinforcements within the network. This involves both short and long duration planned outages that must be evaluated well in advance to ensure adequate visibility for construction planning and outage coordination while satisfying NERC and PJM reliability requirements for N-1 security. As an integral part of operational planning, long-term outage studies, typically conducted several months in advance, are commonly performed on a day-by-day basis using forecasted monthly or seasonal peak load assumptions. This peak load-based approach is widely practiced across the industry to ensure system reliability under worst-case operating conditions. While this approach simplifies analysis, it can be overly conservative for future system studies, as it fails to capture the granular load pattern variability. Consequently, short duration outage requests that may be operationally feasible under more realistic load conditions are often postponed or denied, leading to delays in critical transmission expansion and grid modernization projects.

This paper investigates the operational value of incorporating realistic, multi-granular load forecasting scenarios into long term, contingency-based outage assessments and evaluates their effectiveness relative to conventional peak-based approaches. The granular forecasts are generated using interpretable statistical and Machine Learning models, including SARIMA, Prophet, Gradient Boost, Random Forest etc. followed by bottom-up temporal reconciliation to ensure consistency across daily, weekly, and monthly forecasts. A comparative assessment framework is developed to evaluate the effect of load forecast granularity with consistent study configuration. The comparison evaluates outage feasibility and accommodation, changes in outage decisions, and post-contingency thermal violation severity across multiple forecast granularities. The results demonstrate that incorporating granular load forecasts into assessment reduces unnecessary conservatism and increases outage accommodation, particularly for short-duration requests, without altering the existing reliability criteria. Consequently, the proposed approach can enhance long-term outage coordination, reduce avoidable rescheduling and facilitate the timely execution of transmission reinforcement and grid-modernization projects.

akash.debnath@dominionenergy.com

## 1.1 Introduction

The digital transformation of the U.S. economy, driven by artificial intelligence and cloud computing, has created unprecedented growth in data center load [1]. The “Data Center Alley” in Northern Virginia hosts the largest concentration of data centers in the world, is experiencing particularly significant growth. Dominion Energy, the region's primary Transmission Owner (TO), currently has approximately 70 GW of large-load interconnection requests, of which roughly 25 GW already assigned projected connection dates through 2031. This growth is nearly three times Dominion's historical system peak and requires extensive transmission reinforcement, including new transmission lines, transformers, and substations. Construction of these facilities requires planned outages of existing transmission assets.

Consequently, long-term outage planning has become increasingly challenging for transmission utilities. Each planned outage must be evaluated to ensure that the transmission system remains N-1 secure under credible contingency conditions [2]. NERC and PJM also require coordinated advance assessment of planned maintenance, construction, and reinforcement outages across TOs, making outage scheduling a complex reliability-constrained decision-making problem. For example, outages exceeding 30 days in duration must be submitted to PJM approximately one year ahead of the requested outage window. Although such coordinated advance assessments are essential for maintaining system reliability the effective outage coordination becomes increasingly challenging in rapidly growing load areas because of uncertainties in future transmission system topology, outage schedules, generation availability, and load forecasts.

Existing conservative outage study practices may further restrict outage accommodation in heavily constrained system. Future outage studies are commonly performed on a day-by-day basis within the study horizon using highest expected loading condition for that duration. Although this conservative approach supports reliability assurance, it does not capture the daily and weekly variability of system load and may unnecessarily restrict feasible outages, particularly those of short duration. Industry practices and regulatory requirements increasingly emphasize evaluating system reliability under realistic future load conditions. NERC TPL-001-5.1 requires transmission planning assessments to consider projected real and reactive load forecasts across a broad spectrum of system conditions, while NERC IRO-017-1 & PJM Manual 38 requires coordinated evaluation of transmission and generation outage evaluation, forecasted firm loads, and critical system condition identification [2] [3].

Dominion Energy performs contingency-based outage assessment studies several years advance using its in-house Analysis on Demand (ANODE) platform, which uses the Transmission Analysis and Reliability Application (TARA) and its Outage Reliability Analysis (ORA) module. ANODE integrates EMS-based future network models, transmission outage requests, PJM generation outage schedules, and forecasted load to perform N-1 contingency screening for candidate outages. Each candidate outage is simulated on the future network model and classified as approved, controlled, or rejected based on observed thermal and voltage reliability criteria. The current custom Dominion zonal load forecast is developed internally using state-of-the-art hierarchical forecasting techniques that leverage historical system load from SCADA measurements and system-specific characteristics of the Dominion transmission network.

This study investigates whether coherent daily and weekly load forecasts can improve long-term outage reliability assessments by representing temporal load variability more accurately without altering established reliability criteria. A Python-based forecasting framework evaluates a range of statistical and machine-learning models, selects the best-performing model based on out-of-sample accuracy, and applies temporal reconciliation to ensure consistency across forecast granularities. Using the conventional monthly peak-load approach as the benchmark, the study quantifies the influence of forecast granularity on outage feasibility, accommodation, decision outcomes, and post-contingency thermal violation severity under otherwise consistent study conditions.

The key contributions of this paper are the development of a coherent multi-granular load forecasting framework to support long-term reliability studies and a comparative evaluation of daily, weekly, and monthly forecast granularities for outage reliability studies within Dominion Energy’s existing outage analysis workflow. This evaluation demonstrates how the temporal resolution of load forecasts influences outage reliability assessment outcomes maintaining existing reliability criteria.

The remainder of the paper is organized as follows: Sections II and III formalize the system definition, outage scheduling problem, and evaluation metrics. Section IV describes the proposed forecasting methodology, while Section V presents the ANODE-based case study and comparative assessment. Section VI concludes the paper and outlines directions for future research.

# 2 PRELIMINARIES

Consider a power transmission network, modeled as a topological graph with set of buses $\mathcal{B}$, branch elements (transmission lines and transformers) $\mathcal{E}$ and generators $\mathcal{G}$ . Let $|\mathcal{B}| = B$, $|\mathcal{E}| = L$ and $|\mathcal{G}| =$ G. The complete set of network assets is defined as $\mathcal{A} = \mathcal{B} \cup \mathcal{E} \cup \mathcal{G}$, where each asset $a_j \in \mathcal{A}$, $j = (1,2, \dots, B + L + G)$, represents either a bus, a branch element, or a generating unit. Each asset is assigned to a voltage class through the mapping $v: \mathcal{A} \to \mathcal{V}$ , such that $v(a_j) \in \mathcal{V} = \{500, 230, 115\}$kV, $\forall a_j \in \mathcal{A}$.

## 2.1 Outage Taxonomy

In a study window of $w$ days, let index set of assets that are associated with planned outage requests, $\mathcal{N} \subseteq \{1,2, \dots, B + L + G\}$ . For each asset index $i \in \mathcal{N}$, the corresponding outage record is defined as, $o_i \triangleq (a_i, v(a_i), d_i^s, d_i^e)$, where $a_j \in \mathcal{A}$ denotes the affected asset, $v(a_i)$ denotes the voltage class and $d_i^s, d_i^e$ are the requested start and end date respectively. The start and end days are defined as $1 \leq d_i^s \leq d_i^e \leq w$. For each outage $o_i$, the duration is denoted by $\delta_i = d_i^e - d_i^s + 1$, where $\delta_i$ is measured in days and satisfies $1 \leq \delta_i \leq w$.

## 2.2 Reliability Requirements

For any day $d \in [w] \triangleq [1, 2, \dots, w]$, the transmission system must satisfy the applicable reliability requirements in which a set of outages is active. The resulting network condition must then be evaluated under $N - 1$ contingency conditions. Let, $\mathcal{C}_d$ denote the contingency set and $\mathcal{O}_d$ denote the subset of outages that are active on day $d$. For each contingency $c \in \mathcal{C}_d$, the contingency is then applied on top of the scheduled outage set $\mathcal{O}_d$. The post-contingency AC power flow solution $F_l(\mathcal{O}_d, d, c)$ should converge and must satisfy,

$$|F_l(\mathcal{O}_d, d, c)| \leq F_l^{\max}, \quad \forall l \in [L], \forall c \in \mathcal{C}_d \,, \tag{1}$$

$$V_b^{\min} \leq |V_b(\mathcal{O}_d, d, c)| \leq V_b^{\max}, \quad \forall b \in [B], \forall c \in \mathcal{C}_d \,, \tag{2}$$

Here, $F_l^{\max}$ Is the applicable thermal rating of branch $l$, and $V_b^{\min}$ and $V_b^{\max}$ are the lower and upper acceptable voltage limits at bus $b$, respectively.

## 2.3 Outage Planning Process

Dominion Energy follows a staged reliability-review process, illustrated in Figure 2, that begins with submission of a transmission outage request through iTOA. This is the software solution used by major utilities in USA for outage request scheduling and coordination, switching order, operational logging, event analysis, and reporting. The outage request includes the outage timeline, equipment, clearance requirements, and work description. Submitted outages are incorporated into future reliability studies to identify conflicts before the operating window.

Approximately one to five years ahead, Strategic Planning evaluates major maintenance, construction, and system reinforcement outages against the reliability requirements in eq. (1), (2) using the ANODE-TARA workflow. This early assessment provides visibility into potential reliability limitations and allows sufficient time to revise outage windows, coordinate overlapping work, or develop mitigation strategies. Long-term Planning subsequently evaluates outages from approximately 45 days to one year ahead, while Near-Term Planning performs updated assessments approximately 10 to 45 days before execution.

As the operating date approaches, these assessments incorporate updated topology, outage schedules, load forecasts, and generation availability. In addition to ANODE, planners use the EMS Study Network (ST-NET), which provides current equipment-level connectivity, switching status, and operational topology. From approximately 10 days ahead through real-time operation, the Reliability Engineering desk performs the final validation using the latest outage schedule, forecasted conditions, resource availability, and corrective actions. This staged refinement preserves system reliability while allowing sufficient time to reschedule outages, resolve conflicts, or implement mitigation measures.

## 2.4 Outage Reliability Analysis using ANODE/TARA

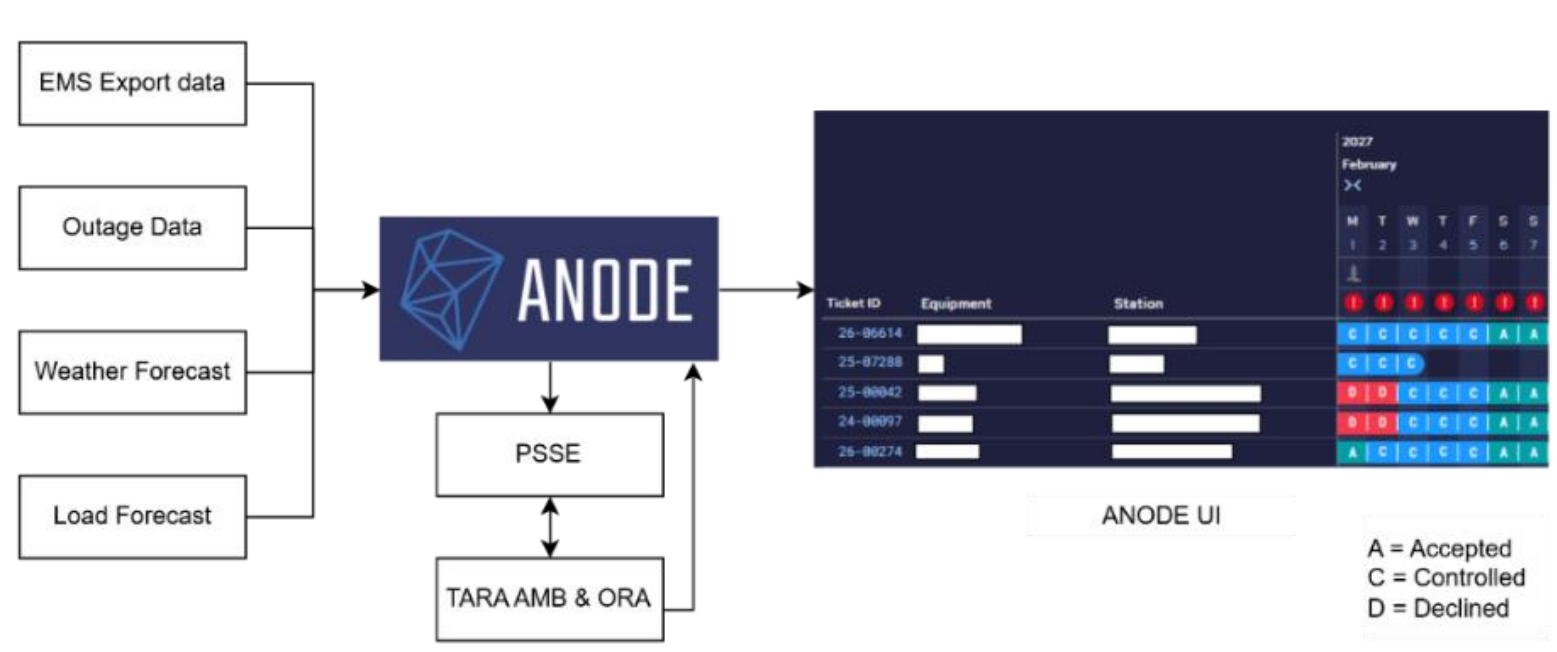

*Figure 1: Outage Planning Module in ANODE*

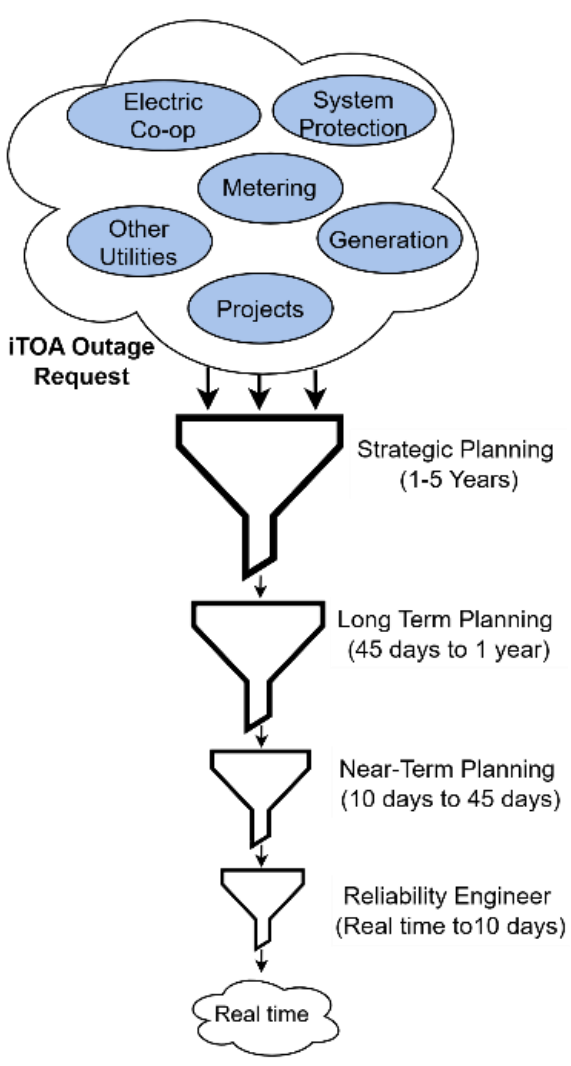

*Figure 2:Outage Study Process*

Analysis on Demand (ANODE) is Dominion Energy's in-house analytics platform, with outage reliability analysis as one of its key capabilities. ANODE functions as an integrated study platform by automatically consolidating and curating electric grid models, weather information, load forecasts, generation dispatch forecasts, and planned transmission and generation outage data into a consistent study environment. ANODE supports automated 10-day, user defined time-period and custom outage studies by preparing inputs, constructing study cases, applying outage and forecast assumptions, initiating the reliability analysis, and organizing results for engineering review.

For each study date, ANODE uses EMS exports and related inputs to construct a PSS/E-compatible base case representing the expected topology. TARA then serves as the study engine. Its Automated Model Builder (AMB) creates the daily case using the scheduled outages, load forecast, and generation conditions, after which the Outage Reliability Analysis (ORA) module performs the assessment.
Under the study outage with all others in base option, all scheduled outages are initially included, and each studied outage is restored individually to determine its reliability impact. This comparison identifies whether an outage introduces or worsens violations, produces significant impacts within acceptable thresholds, or interacts with other outages. ORA first solves the outage-applied AC power flow. A failed solution is classified as non-converged, while a converged case containing monitored thermal or voltage violations is classified as a base-case violation. Only converged cases without base-case violations proceed to N-1 contingency analysis, where ORA compares post-contingency thermal and voltage performance with and without the studied outage. Figure 1 illustrates the outage planning module in ANODE.

An outage is approved when no limiting violation is identified and is retained for subsequent evaluations. If a contingency violation can be mitigated through security-constrained generation redispatch, the outage is classified as controlled; otherwise, it is declined. ANODE then presents the decision, affected facilities, limiting contingencies, load loss, and reliability concerns. Detailed backend branch-flow, voltage, contingency, and outage-impact records allow planners to have further engineering investigation into why an outage was approved, controlled, or declined.

# 3 PROBLEM FORMULATION

Consider a set of planned outages, $\mathcal{N}_w \subseteq \mathcal{N}$requested within the window $w$ days. The objective is to maximize the planned transmission outages accommodation while maintaining compliance with $N-$ 1 reliability requirements mentioned in eq.(1), (2) under forecasted future operating conditions. The network load is supplied through a subset of load buses $\mathcal{L} \subseteq \mathcal{B}$, where the forecasted load at bus $b \in \mathcal{L}$ on day $d$ is denoted by $\hat{y}_{b,d}^{(k)}$. Here, $k \in \mathcal{K}$ represents the forecasting granularity. The first objective is to generate load forecast at each granularity $k \in \mathcal{K}$ based on historical data and the second objective focuses on maximizing outage accommodations subject to multi-granular load forecasting.

## 3.1 Problem Statement

### 3.1.1 Multi-granular Load Forecast

Given a historic time series at $\langle y_{b,t}^{(k)} \rangle_{t=1}^{T}$ observed at each load bus $b \in L$, we compute the total system load time series as, $\langle y_t^{(k)} \rangle_{t=1}^{T} = \sum_{b \in \mathcal{L}} \langle y_{b,t}^{(k)} \rangle_{t=1}^{T}$. Considering the total system load $\langle y_t^{(k)} \rangle_{t=1}^{T}$, we formulate the $H$ step ahead forecasting task as an optimization problem. Let $\hat{y}_{t+h}^{(k)} = f_\theta \left( \langle y_{b,t}^{(k)} \rangle_{t=1}^{T} \right)$, denote the forecast at time $h \in [1, \dots, H]$, where $f_\theta(\cdot)$ is a parametric model with parameters $\theta$. The optimal parameters $\theta^*$ are obtained by minimizing the cumulative loss over the forecast horizon. The forecasting models may operate at different temporal granularities $k \in \mathcal{K}$. Specifically, for $k =$ daily, the forecast remains unchanged. For coarser granularities, a weekly forecast is replicated across all days of the corresponding week, and a monthly forecast is replicated across all days of the corresponding month, yielding the daily forecast sequence $\langle \hat{y}_d^{(k)} \rangle_{d=1}^{w}$ for every $k \in \mathcal{K}$.

### 3.1.2 Maximizing Outage Approval

Since an outage $o_i \in \mathcal{O}_d$ can only be approved when it satisfies the reliability conditions in eq.(1), (2) along with the other scheduled outages in the day, for each active outage-day pair $(o_i, d)$, we define a binary variable $x_{o_i,d} \in \{0,1\}$, such that $x_{o_i,d} = 0$ when the outage $o_i$ does not contribute to voltage and thermal violation in the $N-1$ contingency analysis. The objective is to maximize the total number of approved outages per day over the study window of $w$ days,

$$\max_{|\mathcal{O}_d|} \sum_{d=1}^{w} \sum_{o_i \in \mathcal{O}_d} \mathbb{1}\{x_{o_i,d} = 0\}, \tag{3}$$

subject to the AC power-flow equations and the post-contingency thermal and voltage constraints defined in eq.(1), (2). $\mathbb{1}\{\cdot\}$ is indicator function. The daily system load is determined using $\langle \hat{y}_d^{(k)} \rangle_{d=1}^{w}$, which varies with the selected forecast granularity.

## 3.2 Evaluation Framework

The effects of load-forecast granularity on outage reliability assessment are evaluated using performance metrics that quantify outage approval at the outage-day and full-duration levels. Let, $k \in \mathcal{K} = \{\text{day}, \text{week}, \text{month}\}$ denote the monthly, weekly, and daily load cases, respectively. All approval metrics are evaluated separately for each case $k$.

### 3.2.1 Outage-Day Approval Count (ODAC)

For an outage $o_i$ with requested duration $\delta_i$, ODAC counts the number of days on which the outage doesn't create any violation

$$ODAC(o_i) = \sum_{d=1}^{\delta_i} \mathbb{1}\{x_{o_i,d} = 0\}. \tag{4}$$

### 3.2.2 Outage-Day Approval Rate (ODAR)

This represents how many outage occurrences were accepted on each study date. Let $\mathcal{O}_d$ denote the subset of outages that are active on day $d$. This set is defined as $\mathcal{O}_d = \{o_i \in \mathcal{N}_w | d_i^s \leq d \leq d_i^e\}$. The number of outage occurrences on day $d$ is therefore given by $N_d = |\mathcal{O}_d|$. Therefore, the Outage-Day Approval Rate (ODAR) for day $d$ is then defined as the percentage of active outage that satisfies the reliability criterion on that day $o_i \in \mathcal{O}_d$

$$ODAR(d) = \frac{\sum_{o_i \in \mathcal{O}_d} \mathbb{1}\{x_{o_i,d} = 0\}}{N_d} \times 100\% . \tag{5}$$

$x_{o_i,d} = 0$ indicates that no post-contingency thermal or voltage violations for outage $o_i$ on day $d$.

#### 3.2.3 Acceptance Gain (G)

To quantify the incremental value obtained from granular load forecasting, we define the acceptance gain as the difference in ODAR between two load cases. Let $ODAR_k(d)$ represent the outage-day approval rate on day $d$ under case $k$. The pairwise acceptance gain from case $a$ relative to case $b$ is defined as

$$G_{a|b}(d) = ODAR_a(d) - ODAR_b(d), \qquad a, b \in \mathcal{K} . \tag{6}$$

The evaluated gains are $G_{\text{day|month}}(d), G_{\text{week|month}}(d), G_{\text{day|week}}(d)$, which represent the daily-over-monthly, weekly-over-monthly, and daily-over-weekly acceptance improvements, respectively. Since the gains are computed as differences in ODAR, they are reported in percentage points.

#### 3.2.4 Recovered Outage Count (ROC)

An outage is considered recovered under case $k$ if it is not fully accepted under the monthly case but becomes fully accepted under case $k$. Therefore, ROC is defined as

$$ROC_k = \sum_{o_i \in \mathcal{N}_w} \mathbb{1}\{ODAC_M(o_i) < \delta_i,\ ODAC_k(o_i) = \delta_i\}, \qquad k \in \mathcal{K} . \tag{7}$$

The $ROC_W$ and $ROC_D$ measure the number of outages rejected under monthly peak loading but fully accepted under weekly and daily load representations, respectively.

#### 3.2.5 Weekly Benefit Ratio (WBR)

The average acceptance gain between load cases $a$ and $b$ over the study window $w$ can be defined as:

$$\overline{G}_{a|b} = \frac{1}{w} \sum_{d \in [1,\dots,w]} [ODAR_a(d) - ODAR_b(d)] . \tag{8}$$

Further, weekly benefit ratio (WBR) is defined as

$$WBR_w = \frac{\overline{G}_{\text{week|month}}}{\overline{G}_{\text{day|month}}} . \tag{9}$$

Where, $\overline{G}_{\text{week|month}}$ and $\overline{G}_{\text{day|month}}$ are the average improvements obtained by replacing Monthly loading with Weekly and Daily loading, respectively. A WBR near to one indicates that weekly loading captures most of the benefit achieved by daily loading, whereas a lower value indicates additional value from daily load granularity.

#### 3.2.6 Outage Accommodation Rate (OAR)

Considering a $w$ days study window, $OAR_w$ is defined as the fraction of the requested outages in the window that are approved for their full requested duration, expressed as a percentage

$$OAR_w = \frac{\sum_{o_i \in \mathcal{N}_w} \mathbb{1}\{ODAC(o_i) = \delta_i\}}{|\mathcal{N}_w|} \times 100\% . \tag{10}$$

For short-duration outages, define, $\mathcal{N}_w^S = \{o_i \in \mathcal{N}_w \mid 1 \leq \delta_i \leq 10\}$. Hence, the corresponding short-duration outage accommodation rate is

$$OAR_w(\mathcal{N}_w^S) = \frac{\sum_{o_i \in \mathcal{N}_w^S} \mathbb{1}\{ODAC(o_i) = \delta_i\}}{|\mathcal{N}_w^S|} \times 100\% . \tag{11}$$

#### 3.2.7 Acceptance Opportunity Index (AOI)

The AOI measures the proportion of all studied outage-days that satisfy the reliability assessment. For load case $k$ over study window $w$, the violation burden (VB) and AOI are defined as

$$VB_k^{(w)} = \frac{\sum_{o_i \in \mathcal{N}_w} [\delta_i - ODAC_k(o_i)]}{\sum_{o_i \in \mathcal{N}_w} \delta_i} , \tag{12}$$

$$AOI_k^{(w)} = 1 - VB_k^{(w)} = \frac{\sum_{o_i \in \mathcal{N}_w} ODAC_k(o_i)}{\sum_{o_i \in \mathcal{N}_w} \delta_i}. \quad (13)$$

Unlike OAR, which requires an outage to be approved over its entire duration, AOI credits every accepted outage-day. It captures reduced violation exposure even when an outage cannot be fully accommodated. From an outage planning perspective, a higher AOI indicates lower violation exposure and greater flexibility to schedule, adjust, or coordinate planned outages within the study window.

# 4 PROPOSED METHODOLOGY

## 4.1 Methods for Load Forecasting

The proposed multi-granular load forecasting framework generates base forecasts independently for each forecasting granularity $k$.

### 4.1.1 Time-Series Models

The forecasting task is expressed as learning a mapping from historical load, calendar information, and weather-driven covariates to the future load realization. In an additive representation, the forecasted load is decomposed into a deterministic trend component $\mu_k(t)$, seasonal/calendar components $S_k(C_t)$, autoregressive lag effects $A_k(L_t)$, weather sensitivity $W_k(T_t)$, and unexplained residual term $\varepsilon_t$ [4]. Accordingly, the general forecasting model used in this work can be written as

$$y_t = \mu_k(t) + S_k(C_t) + A_k(L_t) + W_k(T_t) + \varepsilon_t\,. \quad (14)$$

For regression-based machine-learning models, the lag vector is defined as $L_t = \{y_{t-p} : p \in \mathcal{P}\}$ where, $\mathcal{P}$ denotes the set of selected lag orders. These lagged observations allow the model to learn short-term, weekly, and yearly seasonality. The calendar feature vector $C_t$ includes month, day of week, and day of year indicators, enabling the model to capture intra-week, seasonal, and annual operating patterns. The daily peak temperature $T_t$ is included as an exogenous weather covariate because system peak demand is strongly influenced by heating and cooling requirements.

Three classes of forecasting models are considered to produce base forecasts. First, SARIMA models provide a statistical autoregressive formulation in which differencing and seasonal autoregressive-moving-average terms represent persistence and periodic structure in the load series. Second, Prophet represents load as an additive combination of trend, seasonal effects, holiday or calendar effects when available, and residual error, making it suitable for interpretable long-horizon forecasting. Third, machine-learning regression models use lagged load, transformed lag statistics, calendar variables, and daily peak temperature as explanatory features to learn potentially nonlinear relationships between historical operating conditions and future demand.

### 4.1.2 Model Selection and Error Metrics

Each of these models provides a different inductive bias for learning $f_\theta(\cdot)$, and their effectiveness depends on the characteristics of the time series. While SARIMA is specifically effective for stationary series exhibiting strong autocorrelation, models such as Prophet, Holt-Winters, and machine-learning regression models offer greater flexibility in capturing non-stationary behavior and complex seasonal patterns. To ensure a robust and unbiased model selection process, a rolling-origin time-series cross-validation framework is employed on the historical training dataset. The forecasting performance of each model is assessed using a collection of complementary error metrics, including Mean Absolute Percentage Error (MAPE), Symmetric Mean Absolute Percentage Error (SMAPE), Mean Absolute Error (MAE), and Root Mean Squared Error (RMSE). Lower values of these metrics indicate improved forecasting accuracy. The cross-validation results are aggregated across all validation windows, and models are ranked according to their average predictive performance. The highest-performing models are subsequently evaluated on an independent holdout test dataset that was not used during model training or hyperparameter tuning. The final production model is selected based on its ability to consistently minimize forecasting error across both cross-validation and testing stages while maintaining forecast stability and interpretability.

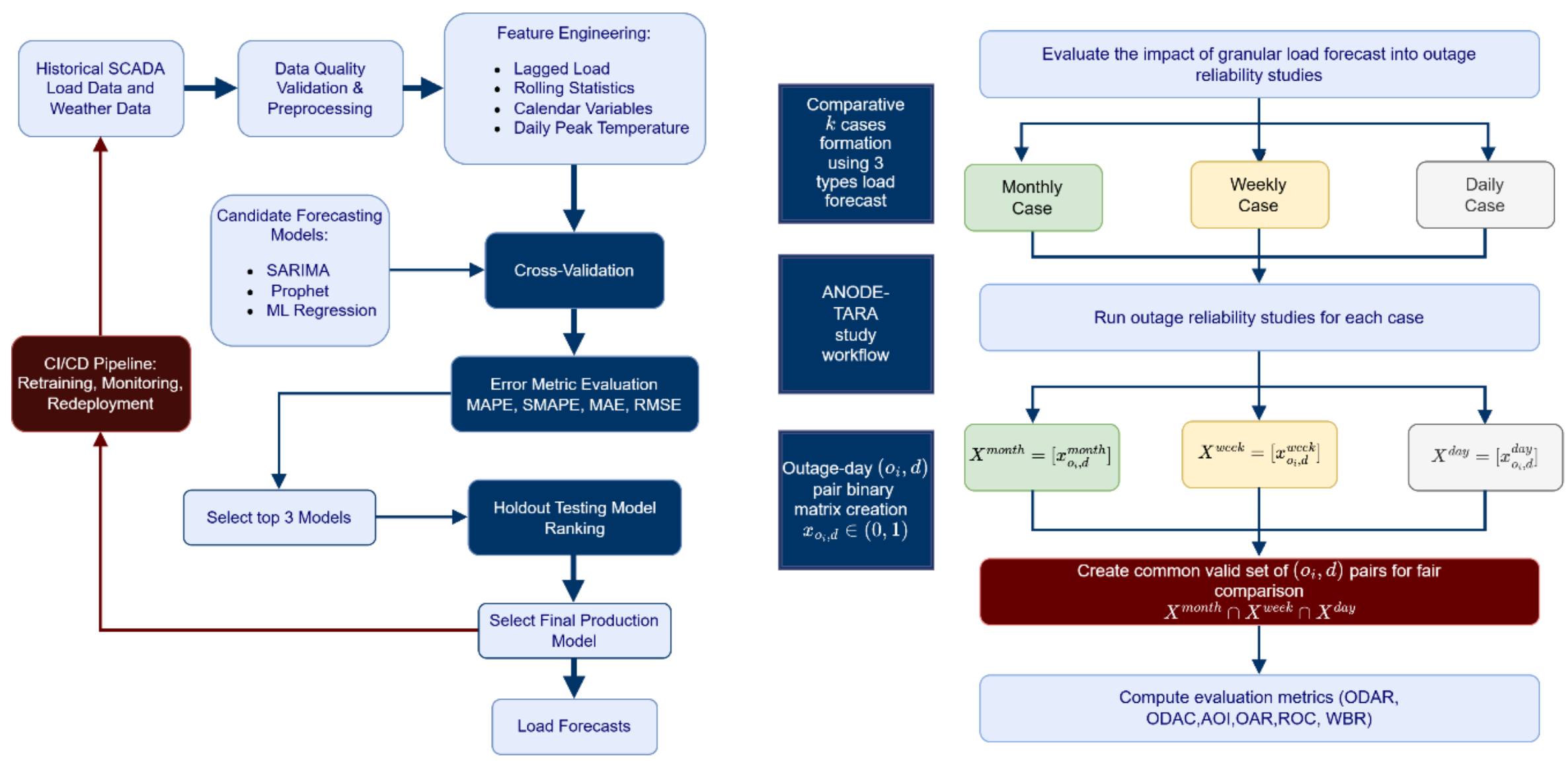

*Figure 3: Load Forecasting Workflow* *Figure 4: Outage Reliability Assessment Framework*

#### 4.1.3 Continuous Integration and Continuous Deployment (CI/CD)

Following model selection, the best-performing forecasting model is deployed within the operational forecasting environment. An automated Continuous Integration and Continuous Deployment (CI/CD) pipeline is implemented to support production forecasting. As new SCADA and weather observations become available, the pipeline performs data quality validation, feature generation, model retraining, performance monitoring, and forecast generation. The forecasting model is periodically re-evaluated against recent observations using the same error metrics, allowing model updates and redeployment when predictive performance degrades. This automated workflow ensures that the forecasting system remains adaptive to evolving load patterns while providing reliable forecasts for long-term outage planning studies. Figure 3 outlines the summary of the load forecasting workflow.

#### 4.1.4 Comparative Outage Reliability Assessment Framework

A comparative assessment framework as depicted in Figure 4 is developed to evaluate the effect of load forecast granularity with consistent study assumptions. For each study window $w$, three cases are constructed using monthly, weekly, and daily load representations. The same network topology, outage request set, contingency definitions and reliability criteria are maintained across the three cases. The load forecast represents the principal variable in the comparison.

Each case is evaluated through the ANODE-TARA outage reliability assessment workflow described in Section II-D. The reliability assessments are executed under the monthly, weekly, and daily loading conditions. The resulting TARA branch-violation extracts are then processed to identify qualifying violations for each active outage-day pair. Dates on which the outage is inactive are not included as studied outage-days. Non-convergences outage-days are excluded consistently across all three cases. The resulting binary matrices are subsequently used to calculate the performance metrics introduced in Section A.3.2

## 5 CASE STUDY

In this section, we discuss the results of using multi-granular load forecasting for outage scheduling studies formalized in Section A.4. The primary objective is to approve more outages with realistic load forecasting scenarios. In Dominion Energy electric transmission network, we model the topology of the network by defining a set of buses $\mathcal{B}$, branch elements (transmission lines and transformers) $\mathcal{E}$ and generators $\mathcal{G}$. For maintaining confidentiality, the exact counts of the line elements kept abstract.

## 5.1 Multi-granular Load Forecasting Results

Like other TOs across the United States, Dominion Energy Transmission has traditionally adopted conservative planning criteria for system upgrades and outage assessments to ensure reliable grid operations. However, the rapid growth of large load customers, particularly data centers, has significantly reduced available transmission capacity margins, making it increasingly difficult to accommodate planned outages. Consequently, reliance on monthly peak load forecasts often results in overly conservative outage assessments that limit operational flexibility. At the same time, developing accurate long-term load forecasts at finer temporal granularities remains challenging due to weather uncertainty. Given that Dominion Energy serves approximately 5 GW of data center load, representing 15%–20% of system demand, exhibit relatively low temperature sensitivity outside the summer months, their substantial magnitude significantly influences loading conditions during outage seasons. This subsection presents the results of the proposed load forecasting methodology outlined in Section 4.1.

### 5.1.1 Data Collection and Sampling

The raw historical data of system load retrieved from the SCADA historian server is typically recorded at intervals of approximately 4 seconds. As the objective of the proposed multi-granular forecasting framework is to support outage planning studies through peak load prediction across multiple temporal granularities $k \in \mathcal{K}$, the high-frequency measurements are aggregated into a daily peak load time series $\langle y_t^{(\text{day})} \rangle_{t=1}^{T}$ spanning over $T = 365 \times 7\ days$. We further partition this dataset in train and test set for evaluating the performance of the forecasting models by selecting a test window of 365 days.

### 5.1.2 Model Selection and Model Evaluation

To establish an appropriate forecasting benchmark, multiple statistical and machine learning forecasting methods are evaluated. The top-performing models identified during cross-validation are subsequently evaluated on an independent holdout testing dataset. Table 1 presents the average forecasting accuracy obtained across the test dataset. Models are ranked according to their average predictive performance. LightGBM model performs best with 42.4% improvement in MAPE score compared to the baseline model. Figure 5 illustrates the train test plot with forecasted time series.

| Model | Category | MAPE | SMAPE | MAE | RMSE | Rank |
|---|---|---|---|---|---|---|
| Seasonal Naive | Baseline | 0.113 | 0.059 | 2145.43 | 2806.62 | 6 |
| SARIMA | Autoregressive | 0.104 | 0.052 | 1959.95 | 2602.84 | 5 |
| Prophet | Additive Decomposition | 0.081 | 0.041 | 1555.30 | 2086.65 | 4 |
| Random Forest | Machine Learning + Exogenous Variables | 0.071 | 0.036 | 1314.25 | 1652.85 | 2 |
| **LightGBM** | **Machine Learning + Exogenous Variables** | **0.065** | **0.033** | **1183.90** | **1516.90** | **1** |
| Linear Regression | Machine Learning + Exogenous Variables | 0.080 | 0.042 | 1483.70 | 1854.61 | 3 |

*Table 1: Holdout Test Performance Comparison*

### 5.1.3 Temporal Reconciliation for Multi-granular Forecasting

Following the generation of forecasts at the daily peak load granularity, the objective is to derive consistent forecasts at weekly and monthly levels. Existing temporal reconciliation approaches are primarily designed for additive hierarchical structures, where higher-level series are obtained by summing observations from lower temporal levels. In the context of peak load forecasting, however, the temporal hierarchy is inherently non-additive. Weekly and monthly peak demands are defined as the maximum load observations within their respective periods rather than the sum of daily peak loads. To address this limitation, a bottom-up reconciliation strategy is adopted. Specifically, forecasts are first generated for the daily peak load series, which serves as the most granular level of the hierarchy. Weekly and monthly peak load forecasts are then obtained by applying the maximum operator to the corresponding daily forecasts within each aggregation period. This approach preserves the physical definition of peak demand while ensuring temporal consistency across forecasting granularities. Figure 3Figure 6 illustrates the resulting daily, weekly, and monthly peak load forecasts obtained using the proposed bottom-up reconciliation framework.

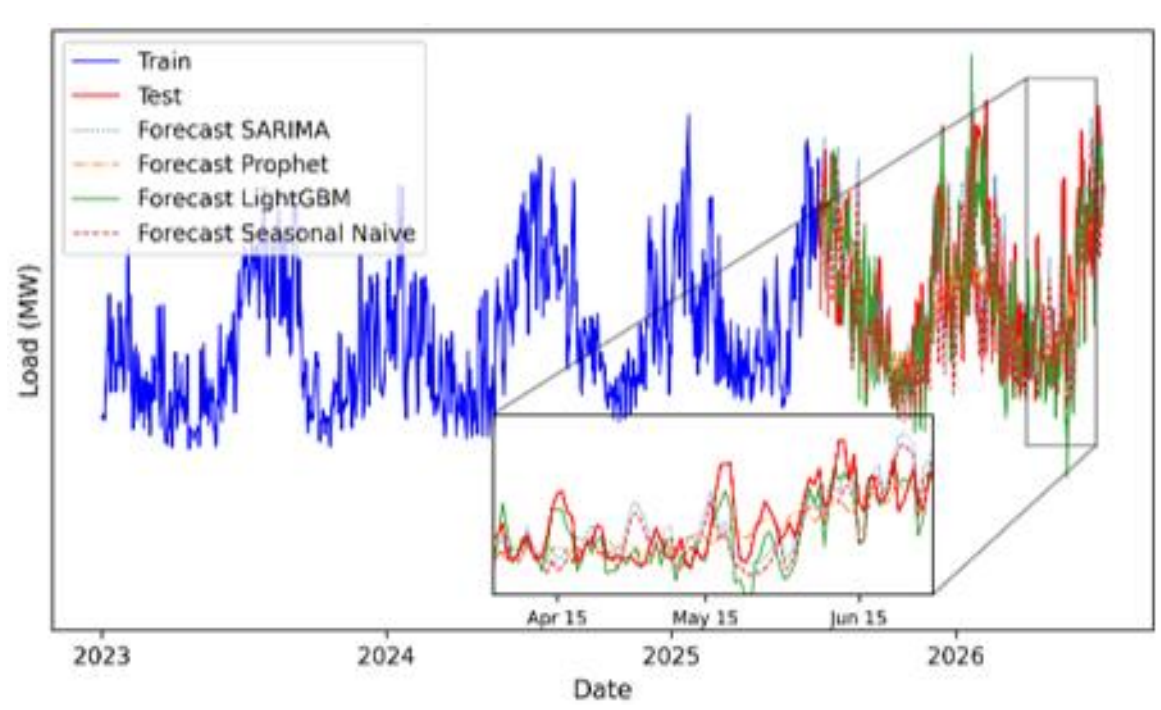

*Figure 5: Test Plot for Dominion System Load Forecast*

*Figure 6: Multi-granular Load Forecast*

## 5.2 Study Configuration

Outage reliability analysis is conducted within Dominion Energy network with $k \in \mathcal{K} = \{\text{day, week, month}\}$ cases over a six-month study horizon, starting from December 2026 through May 2027, yielding 18 study cases. To align with the existing monthly outage-planning process, the horizon is divided into monthly study windows, where $w = days(m)$ $and$ $m \in M$ (set of months in study horizon). Except for the load forecast, the network topology, outage requests, contingencies, and reliability criteria remain identical across all cases. For every month $m \in M$, outage violation outputs are processed separately for each load case $k \in \mathcal{K}$ and converted into an outage-by-date matrix $X^{(k)} = \left(x_{o_i,d}^{(k)}\right)$. Each outage-day pair $(o_i, d)$ is assigned $x_{o_i,d} = 1$ if a violation occurs, 0 otherwise, and NaN if the outage is inactive on that day. Non-convergence outage-day, which populates due to an outage or set of outages, is consistently excluded from all cases to keep the comparison identical. The cases then compared based on the metrics introduced in Section 3.2.

| Month | *ODAR* (%) | | | $G_{\text{day\|month}}$ % | $G_{\text{week\|month}}$ % | $G_{\text{day\|week}}$ % | *WBR* |
|---|---|---|---|---|---|---|---|
| | Month | Week | Day | | | | |
| December | 80.32 | 85.64 | 87.93 | 7.61 | 5.32 | 2.29 | 0.7 |
| January | 69.65 | 82.14 | 90.56 | 20.9 | 12.49 | 8.41 | 0.6 |
| February | 70.22 | 85.12 | 91.28 | 21.05 | 14.89 | 6.16 | 0.71 |
| March | 79.03 | 85.64 | 85.93 | 6.9 | 6.61 | 0.29 | 0.96 |
| April | 73.06 | 78.11 | 81.94 | 8.89 | 5.05 | 3.83 | 0.57 |
| May | 60.84 | 65.25 | 69.25 | 8.41 | 4.41 | 4 | 0.52 |

*Table 2: ODAR & Acceptance Gain over study period*

## 5.3 Results

The results are organized into three complementary subsections to trace the value of forecast granularity from individual outage-day approvals to full-duration outage accommodation and, ultimately, actionable planning opportunities**.** Day-level metrics first quantify changes in violation exposure, after which OAR determines whether those changes are sufficient to accommodate complete outage requests. Finally, AOI and ROC distinguish broader scheduling flexibility from outages that are fully recovered relative to the Monthly case.

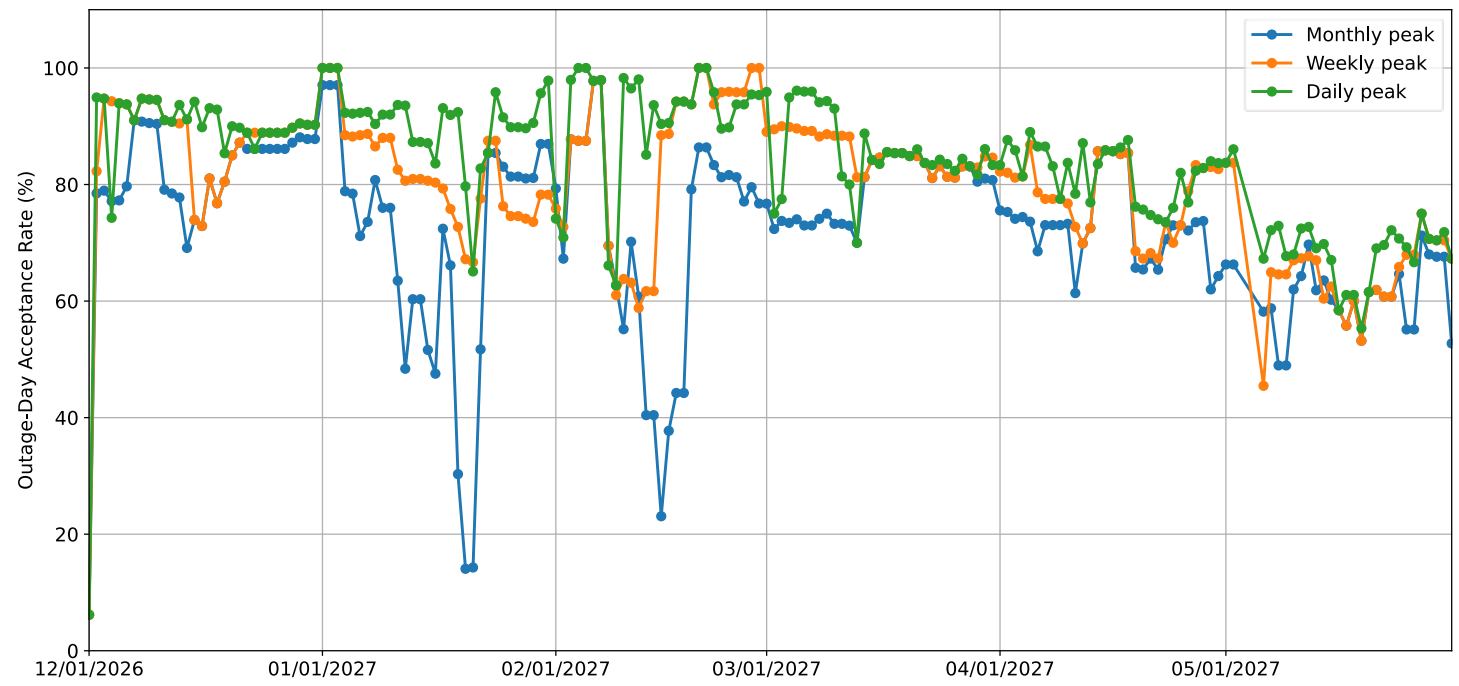

*Figure 7: Day-level ODAR across study horizon*

### 5.3.1 Day level approval performance

Table 3 represents day-level ODAR for month of February 2027. Although isolated dates show zero or negative gains, the overall results demonstrate that granular loading substantially improves outage-day approval compared to monthly case.

shows the daily case generally maintains the highest ODAR across the six-month study period, while both Weekly and Daily loading substantially reduce the approval declines observed under Monthly peak loading. Although Weekly occasionally outperforms Daily on individual dates, the overall trend demonstrates that granular load representation improves outage-day approval, with Daily

| Date | $ODAR$ (%) | | | $G_{day\|month}$ % | $G_{week\|month}$ % | $G_{day\|week}$ % |
|---|---|---|---|---|---|---|
| | Month | Week | Day | | | |
| 2/1/2027 | 79.31 | 75.86 | 74.14 | -5.17 | -3.45 | -1.72 |
| 2/2/2027 | 67.27 | 72.73 | 70.91 | 3.64 | 5.45 | -1.82 |
| 2/3/2027 | 87.76 | 87.76 | 97.96 | 10.2 | 0 | 10.2 |
| 2/4/2027 | 87.5 | 87.5 | 100 | 12.5 | 0 | 12.5 |
| 2/5/2027 | 87.5 | 87.5 | 100 | 12.5 | 0 | 12.5 |
| 2/6/2027 | 97.78 | 97.78 | 97.78 | 0 | 0 | 0 |
| 2/7/2027 | 97.92 | 97.92 | 97.92 | 0 | 0 | 0 |
| 2/8/2027 | 69.49 | 69.49 | 66.1 | -3.39 | 0 | -3.39 |
| 2/9/2027 | 62.71 | 61.02 | 62.71 | 0 | -1.69 | 1.69 |
| 2/10/2027 | 55.17 | 63.79 | 98.28 | 43.1 | 8.62 | 34.48 |
| 2/11/2027 | 70.18 | 63.16 | 96.49 | 26.32 | -7.02 | 33.33 |
| 2/12/2027 | 60.78 | 58.82 | 98.04 | 37.25 | -1.96 | 39.22 |
| 2/13/2027 | 40.43 | 61.7 | 85.11 | 44.68 | 21.28 | 23.4 |
| 2/14/2027 | 40.43 | 61.7 | 93.62 | 53.19 | 21.28 | 31.91 |
| 2/15/2027 | 23.08 | 88.46 | 90.38 | 67.31 | 65.38 | 1.92 |
| 2/16/2027 | 37.74 | 88.68 | 90.57 | 52.83 | 50.94 | 1.89 |
| 2/17/2027 | 44.23 | 94.23 | 94.23 | 50 | 50 | 0 |
| 2/18/2027 | 44.23 | 94.23 | 94.23 | 50 | 50 | 0 |
| 2/19/2027 | 79.17 | 93.75 | 93.75 | 14.58 | 14.58 | 0 |
| 2/20/2027 | 86.36 | 100 | 100 | 13.64 | 13.64 | 0 |
| 2/21/2027 | 86.36 | 100 | 100 | 13.64 | 13.64 | 0 |
| 2/22/2027 | 83.33 | 93.75 | 95.83 | 12.5 | 10.42 | 2.08 |
| 2/23/2027 | 81.25 | 95.83 | 89.58 | 8.33 | 14.58 | -6.25 |
| 2/24/2027 | 81.63 | 95.92 | 89.8 | 8.16 | 14.29 | -6.12 |
| 2/25/2027 | 81.25 | 95.83 | 93.75 | 12.5 | 14.58 | -2.08 |
| 2/26/2027 | 77.08 | 95.83 | 93.75 | 16.67 | 18.75 | -2.08 |
| 2/27/2027 | 79.55 | 100 | 95.45 | 15.91 | 20.45 | -4.55 |
| 2/28/2027 | 76.74 | 100 | 95.35 | 18.6 | 23.26 | -4.65 |
| Average | | | | 21.05 | 14.89 | 6.16 |

*Table 3 : Day-level ODAR & comparative Acceptance Gain*

resolution providing the strongest and most consistent benefit.

As summarized in Table 3 Table 3 : Day-level ODAR & comparative Acceptance Gain daily loading yields the highest average ODAR with the largest $G_{day|month}$ in January and February at 20.90 and 21.05 percentage, respectively. In March, the Daily-over-Weekly gain is only 0.29 percentage points and WBR reaches 0.96, indicating that Weekly loading captures nearly all of the Daily benefit; the lower WBR values in January, April, and May show where Daily resolution adds material value.

#### 5.3.2 Full duration outage accommodation

The outage-duration results in Table 4 further confirms that improved outage-day performance frequently translates into complete outage accommodation. Daily OAR is highest in five of the six months, with particularly large Monthly-to-Daily increases in January, from 21.65% to 64.95%, and May, from 33.85% to 58.46%. March is the exception, where Weekly OAR reaches 75.19% compared with 68.99% for Daily. In Figure 8, short-duration outages (1-10 days) exhibit a stronger granular-load benefit, most notably in May, where acceptance increases from 55.26% under Monthly loading to 76.32% under Weekly and 81.58% under Daily loading. The findings support coarse monthly peak assumptions can disproportionately restrict outages with narrow scheduling windows.

#### 5.3.3 Acceptance Opportunity and Recovered Outages

The Figure 9 and Table 5 shows that granular load representations consistently reduce violation burden across all six study windows. Daily loading produces the highest AOI in every month, with the largest improvements in January and February, where AOI increases from 0.67 to 0.90 and from 0.70 to 0.91, respectively. The nearly identical Weekly and Daily AOI in March indicates that weekly resolution captures almost all available outage-day benefits in that period. A higher AOI indicates that a larger proportion of the evaluated outage-day space is available without reliability violations, which provides greater flexibility to place, adjust, or coordinate planned outages within the study window.

ROC Figure 10 confirms the broader opportunity represented by AOI can be translated into full-duration outage recovery. Daily loading recovers more Monthly-rejected outages than Weekly loading in five of the six months, with the strongest recovery in January and May, whereas Weekly loading performs better in March. Together, AOI and ROC results show that granular forecasts both expand scheduling flexibility and convert part of that flexibility into actionable outage opportunities.

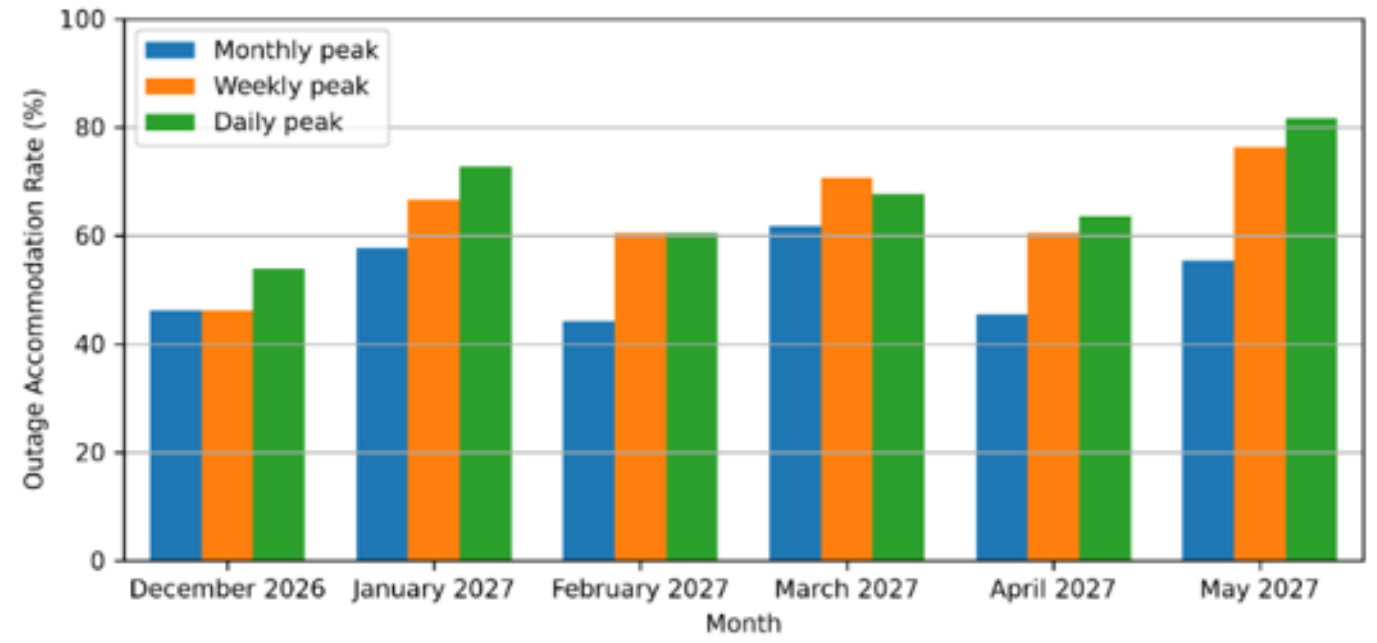

*Figure 8: Short Duration OAR for six month*

Overall, the findings demonstrate that temporally aligned load forecasts improve the fidelity of outage reliability assessment while preserving the same reliability requirements. Weekly forecasts capture much of this value when load conditions are relatively stable, whereas Daily forecasts provide the greatest benefit when day-specific variations meaningfully affect violation exposure and full-duration outage accommodation.

| Month | $OAR_w$ (%) | | | $OAR_w(\mathcal{N}_w^S)$ (%) | | |
|---|---|---|---|---|---|---|
| | Month | Week | Day | Month | Week | Day |
| December | 26.05 | 28.57 | 31.93 | 46.15 | 46.15 | 53.85 |
| January | 21.65 | 56.7 | 64.95 | 57.58 | 66.67 | 72.73 |
| February | 32.65 | 55.1 | 63.27 | 44.19 | 60.47 | 60.47 |
| March | 67.44 | 75.19 | 68.99 | 61.76 | 70.59 | 67.65 |
| April | 50 | 61.64 | 65.75 | 45.45 | 60.61 | 63.64 |
| May | 33.85 | 44.62 | 58.46 | 55.26 | 76.32 | 81.58 |

*Table 4: OAR & Short-duration OAR for study horizon*

| Month | $AOI$ | | |
|---|---|---|---|
| | Month | Week | Day |
| December | 0.79 | 0.85 | 0.87 |
| January | 0.67 | 0.81 | 0.9 |
| February | 0.7 | 0.84 | 0.91 |
| March | 0.79 | 0.86 | 0.86 |
| April | 0.73 | 0.78 | 0.82 |
| May | 0.61 | 0.65 | 0.69 |

*Table 5: AOI*

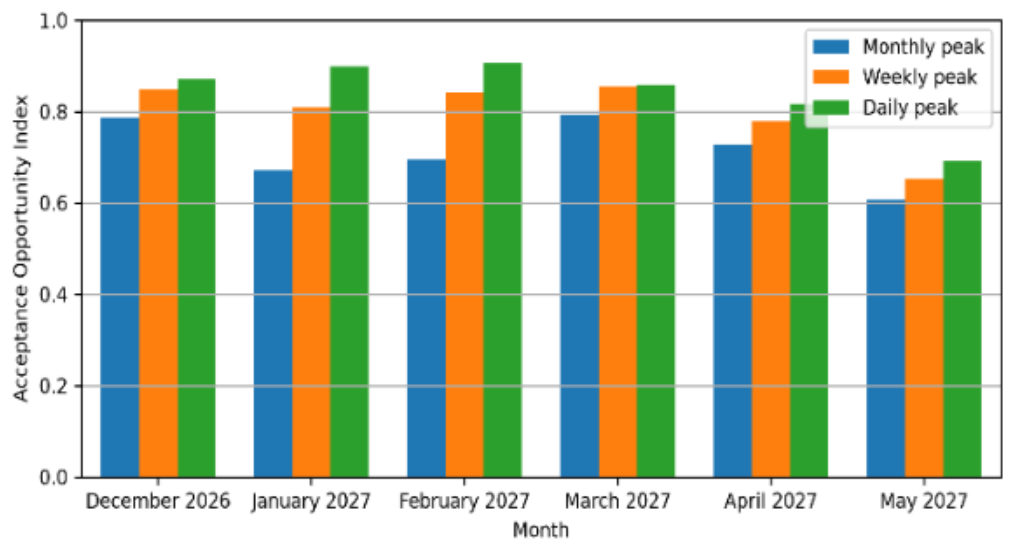

*Figure 9: AOI*

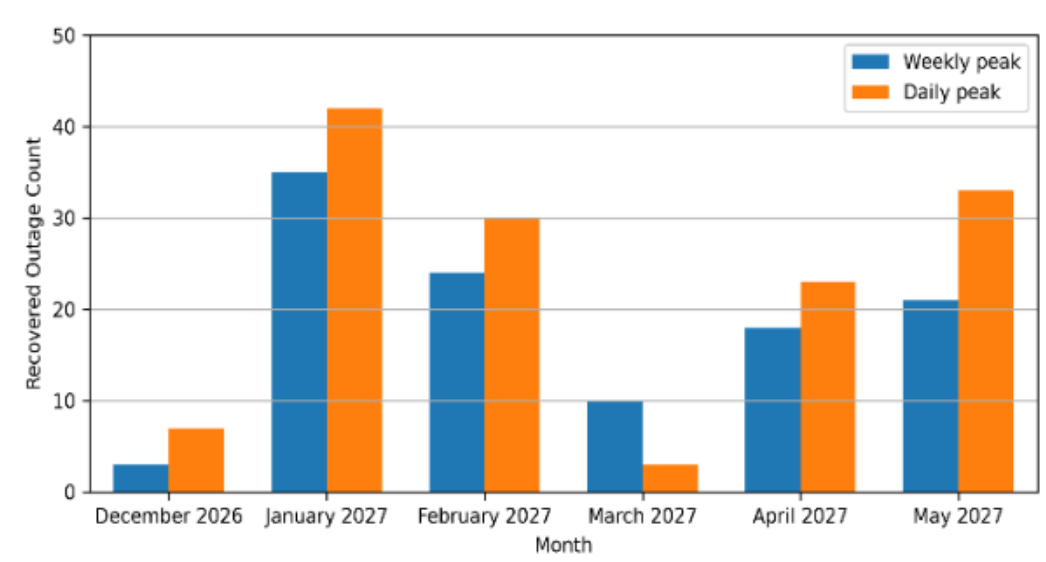

*Figure 10: ROC against monthly case*

# 6 CONCLUSION

This study evaluated the operational value of coherent multi-granular load forecasts in long-term, contingency-based outage reliability assessments. Daily and weekly load representations were compared with the conventional monthly peak load approach under consistent network topology, outage schedules, contingency set, and reliability criteria. The results demonstrate that representing temporal load variability more accurately reduces unnecessary conservatism and improves both outage-day approval and full duration outage accommodation, with particularly notable benefits for short duration outages. Daily loading generally provided the greatest and most consistent improvement, while weekly loading captured a substantial portion of this benefit during periods with relatively stable load conditions. The AOI and ROC results further demonstrate that granular forecasts increase scheduling flexibility and enable additional outages to be fully accommodated. These findings show that load forecast granularity can improve outage planning decisions without altering the underlying reliability requirements. Integrating coherent daily and weekly forecasts into existing outage analysis workflows can therefore support more effective outage coordination, reduce avoidable rescheduling, and facilitate timely transmission reinforcement work. Future work would extend the framework to assess the impacts of granular load forecast on transmission transfer capacity and incorporate forecast uncertainty through probabilistic scenario.